%% file: paper.tex
\documentclass[sigconf]{acmart} 
\pdfoutput=1
\renewcommand\footnotetextcopyrightpermission[1]{} 

\usepackage{booktabs} 

\usepackage{amsfonts}
\usepackage{amsthm}
\usepackage{mathtools}
\usepackage{subcaption}
\usepackage{xcolor}
\graphicspath{{./figs/}}

\DeclarePairedDelimiter{\ceil}{\lceil}{\rceil}

\newcommand{\rank}[1]{$#1^{\text{th}}$}

\newcommand{\Levy}{L\'{e}vy }

\newcommand{\LWNoSpace}{\Levy Walk}
\newcommand{\RWNoSpace}{Random Walk}
\newcommand{\RWPNoSpace}{Random Way-point}
\newcommand{\MRWPNoSpace}{Manhattan Random Way-point}

\newcommand{\LW}{\LWNoSpace { }}
\newcommand{\RW}{\RWNoSpace { }}
\newcommand{\RWP}{\RWPNoSpace { }}
\newcommand{\MRWP}{\MRWPNoSpace { }}

\newcommand{\ManhattanRandomWayPoint}{\MRWP}

\begin{document}
\title[Improved Bounds on Manhattan Random Waypoint Model]{Improved Bounds on Information Dissemination by Manhattan Random Waypoint Model}

\author{Aria Rezaei}
\affiliation{%
  \institution{Stony Brook University}
}
\email{arezaei@cs.stonybrook.edu}


\author{Jie Gao}
\affiliation{%
  \institution{Stony Brook University}
}
\email{jgao@cs.stonybrook.edu}

\author{Jeff M. Phillips}
\affiliation{%
  \institution{University of Utah}
}
\email{jeffp@cs.utah.edu}

\author{Csaba D. T\'oth}
\affiliation{%
  \institution{Cal State Univ. Northridge}
}
\email{csaba.toth@csun.edu}

\renewcommand{\shortauthors}{A.~Rezaei et al.}

\begin{abstract}
\input{000-abstract.tex}
\end{abstract}

%

\maketitle

\begin{abstract}
\end{abstract}

\section{Introduction}\label{sec:intro}
\input{010-introduction.tex}

\section{Background and Related Works}\label{sec:rel-work}
\input{020-related-work.tex}

\section{Bounds on Flood Time}\label{sec:bounds}
\input{030-contribution.tex}

\section{Experiments}
In this section, through a series of experiments we test the accuracy of our discovered bound.
First, we test our model where agents are moving in a torus-like grid, following \MRWP model.
Next, using bike sharing records in 3 major cities, we create synthetic trajectories in
real-world road networks, and compare simulated behavior of flood time against our model.
Finally, using GPS traces of taxis in a major city, we verify our model against a real-world case of information dissemination via mobile agents.
\input{041-simulation.tex}
\subsection{Real-World Data}\label{sec:real-data}
\input{042-real.tex}
\section{Towards Bounding the \LW}\label{sec:levy}
\input{050-levy.tex}
\section{Conclusion}\label{sec:conclusion}
\input{060-conclusion.tex}

\section*{Acknowledgements}
A. Rezaei and J. Gao 
acknowledge support through NSF CCF-1535900, CNS-1618391, and DMS-1737812. J. Phillips 
acknowledges support by NSF CCF-1350888, ACI-1443046, CNS-1514520, and CNS-1564287.
Research by C.D.~T\'oth was supported in part by NSF CCF-1422311 and CCF-1423615.
The experiments were conducted with equipment purchased through NSF CISE
Research Infrastructure Grant No. 1405641.
The authors 
thank Dagstuhl Seminar 15111 on Computational Geometry 
during which some of the ideas were developed. 

\bibliographystyle{ACM-Reference-Format}
\bibliography{paper,jiepub,privacy,jie-4}

\end{document}

%% file: 000-abstract.tex
With the popularity of portable wireless devices it is important to model and predict how information or contagions spread by natural human mobility -- for understanding the spreading of deadly infectious diseases and for improving delay tolerant communication schemes. Formally, we model this problem by considering $M$ moving agents, where each agent initially carries a \emph{distinct} bit of information. When two agents are at the same location or in close proximity to one another, they share all their information with each other. We would like to know the time it takes until all bits of information reach all agents, called the \textit{flood time}, and how it depends on the way agents move, the size and shape of the network and the number of agents moving in the network.

We provide rigorous analysis for the \MRWP model (which takes paths with minimum number of turns), a convenient model used previously to analyze mobile agents, and find that with high probability the flood time is bounded by $O\big(N\log M\ceil{(N/M) \log(NM)}\big)$, where $M$ agents move on an $N\times N$ grid.
In addition to extensive simulations, we use a data set of taxi trajectories to show that our method can successfully predict flood times in both experimental settings and the real world.

%% file: 010-introduction.tex
It has always been an interesting research topic to understand human mobility and how contagions spread via such motion. One of the motivations is to understand how infectious diseases spread by moving agents. Think of  
a strain of an infectious virus
such as SARS or Ebola. When two individuals are at the same location at the same time, there is a possibility for one to spread that strain of virus to the other. Therefore the spread of contagions in a population is highly dependent on the density of the population and how individuals move around. In another example, human motion can be used for our benefit. It has become common that people carry wireless devices around.  Short range low cost wireless communication can be established at the contact events to allow energy efficient information exchange.  In this case the mobility model influences how 
long it takes for a piece of information from one node to reach all nodes in the network.

There have been two main approaches to study human mobility in the literature: data-driven methods versus theoretical analysis.
In recent years wireless technology has made it possible to collect a large amount of mobility data through wireless devices. There has been a lot of work on finding exciting patterns in human movements and information spreading in real-world data~\cite{JiaxinVehicleNetwork,H1N1Spread1,H1N1Spread2}. It has been shown that human mobility is immensely complex.
Accurate models can be built using historic traffic data to predict agents' locations and social ties~\cite{song2010limits,Eagle08092009}. But these models each work only for a specific scenario. It is unclear how these models can help us analyze asymptotic behavior of moving agents or whether a model generalizes to a different geographical location, a different travel modality, or a different group of people. 
Furthermore, long-term mobility data can be identity revealing. Even with great efforts to anonymize the data and with removing big fractions of it, individuals are identifiable by their movement patterns. A seminal work revealed that $4$ randomly selected points in an hourly location sequence of a person recorded over $15$ months via cellphone antennas is enough to make that person identifiable among $1.5$ million individuals \cite{MovementPrivacy::UniqueInCrowd}. As a result, mobility data sets are usually not published by companies due to concerns over user privacy, except for a few special cases of shared vehicles (taxis or shared bikes). 

On the other hand, an extensive amount of work has been dedicated to mathematical models of mobility and their asymptotic behaviors.
Although these models cannot compete with the accuracy of data-driven models in the presence of enough historic mobility data, they have been used for their rigorous analysis and their ability to predict future events with provable certainty. Over the years theoretical models have evolved from simplistic models, inspired by known physical phenomena in real world, to more sophisticated ones, taking into account the complexity of human mobility.
We briefly review these models and their analytical results below.
For a comprehensive survey on
these models refer to \cite{MovementModelSurvey}.

\begin{itemize}

\item \textbf{Random Walk:} Perhaps the most studied movement model. Inspired by the movement of floating particles in a liquid or gas, called \emph{Brownian
Motion} \cite{FeymanLecturesBrownianMovement}, in its simplest version
an agent starts its movement in an arbitrary node in a given network. At each time step, the agent chooses one of the neighbors of its current node \emph{uniformly at random} and moves to that neighbor.  There are variations in which agents can rest in their current position for a period of time or use a non-uniform transition probability when choosing a neighbor.

\item \textbf{Random Direction Model:} In this model, an agent chooses a random direction and
possibly a random velocity, then moves in that direction until it collides with the boundary of
the network. The agent then chooses a new direction and velocity and continues as before.

\item \textbf{\RWPNoSpace}: In this model, an agent, starting from an initial position
in the network, chooses the next destination from \emph{all} nodes in the network \emph{uniformly at random}. Then, using one of the shortest paths, moves towards the destination and after reaching it finds a new destination with the same method. This has been widely used in modeling human motion and in many prior simulations for mobile networks~\cite{Hu2000-su,Johnson1996-ib,Perkins2001-ij,Broch1998-ip,Boudec2006-ok}.

\item \textbf{\ManhattanRandomWayPoint:} This is a special case of the {\RWPNoSpace}   model~\cite{Boudec2006-ok}. In a grid
(or \emph{torus}) networks, agents move to the destination with as few turns as possible. Thus they travel first horizontally and then vertically (or vice versa) to the destination.
This model is inspired by the fact that in urban streets, turning can be
time-consuming~\cite{Crescenzi2009-mp,Clementi2011-dn}. 

\item \textbf{L\'evy Walk:} Studies on intelligent moving agents, especially humans, have
revealed that the distance to the next destination chosen by such agents
seems to follow a \emph{fat-tailed} distribution~\cite{HumansMoveFatTailed,OriginBurstyHumanMovement,Viswanathan2002-yf}.
A popular movement model in this category is called \emph{\LW}~ \cite{LevyFlight}.
This model is similar to \RWPNoSpace, but instead of choosing a new destination uniformly
at random, an agent chooses a node as its next destination with probability proportional to
the inverse of its distance from the agent's current position, to some power $\alpha>0$.

\end{itemize}
%


\subsection{Our Contributions}\label{subsec:contrib}
In this paper we provide improved upper bounds on the rate of information dissemination when agents move according to the \MRWP model. Through a series of simulations we show that, combined with bounds on \RWNoSpace, our bounds lead to a new conjecture on the time it takes for information to disseminate through a network when agents' movements follow the \LW~model, a challenging question that remains widely open. Finally, we report the result of a series of experiments we have designed, which show that our model is capable of predicting trends on experimental settings, as well as real-world data.

We formally define our problem as follows.
Consider a set of $M$ autonomous agents,
each starting at time 0 with a unique bit of information, $b_i$, at a node selected uniformly at random
in an $N \times N$ torus\footnote{A grid where nodes in the boundaries are each connected to their corresponding node in the opposite boundary.}
denoted by $G$. Agents all follow the same movement model and they share information
with each other when they meet during their move. Meeting is defined as
being at the same location at the same time, where the location can be inside a node
or on an edge between two nodes. Agents start their movement at the beginning of each
time step in a synchronized fashion. We consider uniform speed for all agents\footnote{In some previous definitions of \RWP the agents choose their speed uniformly at random from a range. But this choice will lead to the average moving speed to be decreasing over time~\cite{Yoon2003-hb}. Also in reality vehicles/pedestrians often move with a fixed speed.} and transmission
radius is practically $0$ as agents have to be collocated in order to pass along information, for simplicity.
However our findings can be extended for arbitrary constant transmission radius.

In the above setting, we are interested in finding bounds on the time it takes until \emph{every} agent finds out about \emph{every} piece of information. This value is called the \emph{flood time} ($T_F$).
An equally important statistic is the time it takes for all agents to learn a
\emph{specific} bit of information, called \emph{broadcast time} and denoted by $T_B$. Clearly $T_B \leq T_F$. Using the \emph{union bound}, any upper bound on $T_B$ extends to $T_F$, too, if the probability of the bound occurring is \emph{sufficiently high}\footnote{If $\Pr\{T_B > T\} < 1/N^d$, for some $d>0$, we know that $\Pr\{T_F > T\} < M/N^d < 1/N^{d'}$, when $d$ is sufficiently large.}.
Both $T_F$ and $T_B$ are important statistics in various applications.  In the case of disease spreading, $T_B$ corresponds to the time when an infection of \emph{any} agent would have been
passed to the entire population.  In a delay-tolerant wireless mobile network, 
$T_F$ corresponds to the time in the past, from which we can assume all information has been shared across the network; we can predicate the start of a new protocol based
on assuming all agents are up to date after this delay.  In mobile social
networks, $T_B$ corresponds to the time it takes a new piece of information to
permeate society. Overall, both $T_F$ and $T_B$ are important statistics which capture
information flow in a network, and will be the focus of our study.

Our contributions are as follows:
\begin{itemize}
\item We find a new upper bound for $T_F$ and $T_B$ that is tight for a wide range of settings.
Specifically, when $M$ agents move on an $N \times N$ grid with torus topology, we show
$T_B \leq T_F = O(N \log M \lceil \frac{N}{M} \log(NM)\rceil)$.
This bound improves upon recent upper bounds for topologies with more complex boundary conditions.
\item We analyze the relation between \RWNoSpace, \MRWP and \LWNoSpace. Through simulations, we  show empirically that the \LW model can be understood by carefully interpolating
between the heavily studied Random Walk model and our new results on the \MRWP model.
\item We validate the theoretical bounds in a number of empirical studies using simulated scenarios, bike and taxi trajectory data sets.
\end{itemize}

%% file: 020-related-work.tex
Since the advent of social networks, researchers studied how information (e.g., rumors or viral videos) spreads through a network \cite{StaticKleinberg,StaticLeskovec,StaticKempe}. The rate of change in these networks is
slow enough that they can be considered \emph{static} throughout the
course of information dissemination. This assumption simplifies the mathematical models tremendously.
These studies do not fit \emph{dynamic} mobile networks, due to the high rate of topology changes. The spreading behavior heavily depends on how agents move
\cite{KleinbergWirelessEpidemic}.

For analyzing the mobility models mentioned earlier, arguably the simplest model for a geographically spread network is a 2D grid. In some metropolitan settings the downtown area is a reasonable grid. Wrapping the grid around to a torus has been commonly adopted in prior papers that analyze information spreading in a mobile network. The benefit of the torus model is to get rid of the boundary effect, thus simplifying the analysis. The torus also removes the boundary effect in the \RWP model, which often caused unwanted artifacts in simulations~\cite{Chu2002OnTA,Bettstetter2001-jr,Royer2001-ra}. 

For direct comparison, we review prior results in the same format as ours, where $N$ and $M$ are decoupled and there are no assumptions on their ratio, near-zero transmission radius ($R \approx 0$), and where each \emph{move} consists of walking along a path from source node to destination node, rather than \emph{jumping} to the destination instantaneously.


The \RW movement model has been extensively studied and tight bounds on many characteristics have been fully resolved~\cite{RandomWalkAncient1,RandomWalkAncient2,RandomWalkBook}.
One of the tightest bounds for \RW is given by Patterin et al. \cite{RandomWalkBound}. They found that
even with a very small transmission radius, broadcast time does not depend on the relation between
the mobility speed and the transmission radius. They prove that with high probability (w.h.p):
\begin{equation}\label{eq:random_walk_bound}
T_B = \tilde{O}\Big(N\ceil[\Big]{\frac{N}{\sqrt{M}}}\Big).
\end{equation}
Keeping track of a bit of information $b$, they first divide the initial grid into smaller cells
and find the time by which an arbitrary cell is \emph{infiltrated} by an agent carrying $b$.
Then they show that this infiltrating agent will inform the majority of agents near this cell and spread the information \emph{locally}.
After the local spread is done within the cell, information \emph{leaks} into adjacent cells and this whole process is repeated.
Ultimately, every cell is infiltrated and every agent in the network finds out about $b$.

Clementi et al.~\cite{ManhattanFlood} have proved bounds for $T_F$ in Manhattan-like grids where agents move according to the \MRWP model.
Their setting slightly differs from ours as they do not employ boundary loops, which results in two zones with very different traffic of agents. On the one hand, nodes in the \emph{central zone} are visited by agents across all nodes in the grid with high probability. On the other hand, the periphery, called \emph{suburb} areas, are starved of agents: the probability that an agent passes through these areas is significantly lower than that of the central zone.
They prove that w.h.p. $T_F = \tilde{O}\big(N/R + N^3/vR^2M\big)$, where $R$ is the transmission radius and $v$ is the agents' speed.
We can rewrite this in our setting as:
\begin{equation}\label{eq:bound_manhattan}
T_F = \tilde{O}\Big(N + \frac{N^3}{M}\Big).
\end{equation}
They found that the bulk of the Flooding Time is devoted to
carrying the information to the ``suburbs,'' as it requires a flow of informed agents traveling
from the central zone to the suburbs.


A different line of work focused on solving the problem in a general platform, oblivious to geometric considerations. Clementi et al.~\cite{MarkovBound} have derived an upper bound for $T_F$, and the 
\emph{mixing time}\footnote{Mixing time is the time needed for a Markov chain to reach its stationary distribution, starting from an arbitrary distribution.} of the Markov chain
corresponding to agent movements, as well as how independent the collisions between different pairs are, play a major role in the analysis of $T_F$.
Applying \MRWP model to their general bound yields the following:
\begin{equation*}
T_F = O\Big(\frac{N}{v_{\rm max}}\big(\frac{N^2}{MR^2} + 1\big)\log^2{M}\Big),
\end{equation*}
where $v_{\rm max}$ is the maximum speed of any agent. 
Rewritten in our setting, the bound is:
\begin{equation}\label{eq:markov_rwp}
T_F = \tilde{O}\Big(N\ceil[\Big]{\frac{N^2}{M}}\Big).
\end{equation}

Since their method is very general, their bound is not competitive with the bounds on specific movement models and networks.
Table~\ref{table:all_bounds} shows that our bound is a significant improvement over these bounds on different movement models. 
\begin{table}[!tbh]
\centering
\caption{Recent bounds on \MRWP (MRWP), \RWP (RWP) and \RW (RW).}
\label{table:all_bounds}
\begin{tabular}{|c | c | c|}
 \hline
 \textbf{Authors} & \textbf{Model} & \textbf{Bound} \\
 \hline
 \textbf{Ours} & \textbf{MRWP} & \textbf{$\tilde{O}(N\ceil{N/M})$} \\
 Clementi et al. 2010~\cite{ManhattanFlood} & MRWP & $\tilde{O}(N + N^3/M)$ \\
 Clementi et al. 2015~\cite{MarkovBound} & RWP & $\tilde{O}(N\ceil{N^2/M})$ \\
 Pettarin et al. 2011~\cite{RandomWalkBound} & RW & $\tilde{O}(N\ceil{N/\sqrt{M}})$ \\
 \hline
\end{tabular}
\end{table} 

Rigorous analysis of the \LW model is much more challenging due to the strong spatiotemporal correlation~\cite{Birand2011-mx,Shinki2017-sv,Lee2011-jg}. 
The most relevant work is done by Wang et al.~\cite{Wang2014-nj} where the authors have analyzed the distribution
of the minimum time needed until a piece of information reaches a certain region.
However, to the best of our knowledge, there are no bounds for $T_F$ when the agents are moving according
to a \LWNoSpace.

Besides theoretical work, there has been a lot of empirical analysis of how information can spread through opportunistic peer communication among mobile agents using simulations or empirical evaluations. Protocols for reducing communication cost (e.g., to avoid a message be delivered to a node multiple times) have been studied extensively~\cite{chu2002scalable, intanagonwiwat2000directed, peng2000reduction}.
Last, there has also been work on a model assuming that a supporting static wireless network is in place, which helps to cache and propagate these events~\cite{zhou09opportunistic}. This model is different from ours. 

%% file: 030-contribution.tex
In this section, we present and prove our main theoretical results on the flooding time on torus networks. We start with a trivial lower bound.

\begin{theorem}\label{theorem:lb}
For $M$ agents initially positioned at uniformly random nodes and moving
with constant speed in an $N \times N$ torus, with constant probability, we have:
\begin{equation}\label{eq:triv_lb}
T_F \geq T_B \geq \Omega(N).
\end{equation}
\end{theorem}
\begin{proof}
Let $b_i$ be the bit of information initially carried by agent $A_i$, and
$A_j$ the farthest agent to $A_i$ at time $0$. The time it takes until $b_i$
reaches $A_j$, denoted by $T(A_j, b_i)$, is a lower bound for both $T_F$ and $T_B$.
Since we are assuming uniformly random initial positions, with $1/2$
probability the distance between $A_i$ and $A_j$ is at least $N/2$.
Also note that information can only travel as fast as agents can. This means that
at each time step the distance between $A_j$ and the closest copy of $b_i$
is reduced by at most 2 units. As a result, with probability $1/2$, $T(A_j, b_i)$ is at least
$N/4$, which completes the proof.
\end{proof}

Now we move forward to our main theorem on the upper bound for \MRWP  model.

\begin{theorem}\label{theorem:main}
For $M$ agents moving according to the \MRWP model with constant speed in an $N \times N$ torus,
with high probability\footnote{High probability in our work means at least $1 - 1/N^d$ for some constant $d > 0$.}, we have:
\begin{equation}\label{eq:main_eq}
T_B \leq T_F = O\Big(N\log{M}\ceil[\Big]{\frac{N}{M}\log{(NM)}}\Big).
\end{equation}
\end{theorem}

W present the proof in two parts. The first part is to analyze when and with what probability two agents have a collocation event so they can share information. This involves a few necessary conditions: time-wise the moves made by two agents with a collocation event need to overlap; geometrically their trajectories need to have an intersection; and thirdly, they arrive at the intersection at the same time.  Next we need to analyze the global property: how information sharing enabled by collocation events leads to global dissemination. The next two subsections focus on these two parts of the proof respectively.

\subsection{Bounding Collocation Probability}
\input{031-collocation.tex}

\subsection{Bounding the Flood Time}
\input{032-flood.tex}

%% file: 031-collocation.tex
We consider agents moving non-stop following  \MRWP model, and partition the mobility trace of each agent into \emph{moves} between randomly chosen destinations. First, observe that any move by an agent following \MRWP model takes $\Omega(N)$ time with constant probability.  We call the two straight parts of a movement, one horizontal and the other vertical, \emph{segments}. Note that any move in this model can have at most two segments of
different directions appearing in an arbitrary order. 

We say two agents have a \emph{connection} during their moves if they are at the same location at the same time. For that to happen, the moves of the two agents must at least overlap over time. 

\begin{definition}
\textbf{\textit{Strongly overlapping moves}} are moves made by two agents where
the \emph{time interval} of a segment of an agent's move is completely contained within
the time interval of the other agent's whole move.
\end{definition}

Note that this overlap only needs to happen in the time interval of two moves
and no condition is imposed on their geometric locations.

\begin{lemma}\label{lem:strong-overlap}
Every move  $M_i$ of an agent $A_i$ strongly overlaps with at least one of the moves of
another agent $A_j$, say $M_j$.
The starting moment of the moves can be at most
$N$ time-steps apart and 
with constant probability, a segment of $M_i$ will have a time duration overlap of $\Omega(N)$ with a segment of
$M_j$.
\end{lemma}
\begin{proof}
First consider the move $M^-_j$ of $A_j$ that is still active when $M_i$ starts.
If $M^-_j$ ends after the first segment of $M_i$, then $M^-_j$ covers the first segment of $M_i$ and the two moves strongly overlap (Figure~\ref{fig:lem-overlap} case (i)).
If not, then consider the next move of $A_j$, $M^+_j$. If $M^+_j$ ends after
$M_i$ ends, $M^+_j$ completely covers $M_i$'s second segment and the two moves strongly overlap (Figure~\ref{fig:lem-overlap} case (ii)).
If not, then $M^+_j$ must be completely covered by $M_i$ and the two moves strongly overlap (Figure~\ref{fig:lem-overlap} case (iii)).
We have shown that a strong overlap occurs. Since the time duration of each move can be at most $N$, the starting point of $M_i$ is
at most $N$ time-steps apart from the starting points of both $M^-_j$ and $M^+_j$.

\begin{figure}[tbh!]
\centering
\vspace*{-2mm}
\includegraphics[width=.55\linewidth]{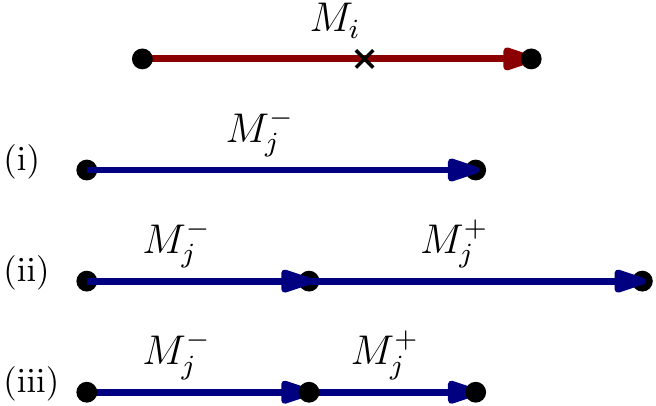}
\caption{Any move of one agent strongly overlaps with at least one of the moves of another agent.}
\label{fig:lem-overlap}
\end{figure}

Without loss of generality, assume a segment $S_j$ of $M_j$ is covered completely by $M_i$.
With probability $1/2$, the time duration of $S_j$ (equal to its length due to the constant speed assumption)
is at least $N/4$. Now take the segment of $M_i$
with the longest time duration overlap with $S_j$, and denote it by $S_i$. This overlap should be at least
$|S_j|/2$. As a result, with probability $1/2$, the time duration of this overlap between $S_i$ and $S_j$
is at least $N/8$, as required.
%
%
\end{proof}

\begin{figure}[tbh!]
\centering
\includegraphics[width=.85\linewidth]{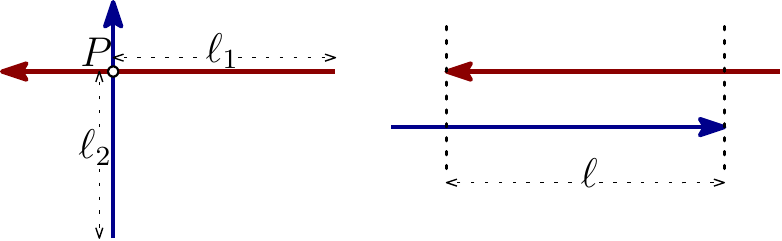}
\caption{The segments are trimmed to duration of their overlap, $\ell$. Left: The collocation event of two agents happens if the distance between the intersection point ($P$)
and the starting point of both segments is equal ($\ell_1 = \ell_2$). Right: Connection
happens if there is a non-zero overlap between two segments and there are $2\ell$ such placements for the blue segment if the red segment is fixed.}
\label{fig:connection}
\end{figure}

\begin{lemma}\label{lem:l/n^2-connection}
If two agents $A_i$ and $A_j$ have two segments $S_i$ and $S_j$ with a time interval overlap of $\ell$,
they meet with probability $\Theta(\ell/N^2)$.
\end{lemma}

\begin{proof}
We trim both segments to the duration of their overlap, making them of equal length $\ell$.
All cases of a connection between the two agents can be reduced to three main cases below by rotating
the torus or swapping the agents:

\begin{enumerate}
\item $S_i$ is horizontal and $S_j$ vertical.
Here, the two agents \emph{connect}, if the two segments intersect geometrically at a point $P$ and
$P$ is at equal distance from the starting points of
$S_i$ and $S_j$ (see Figure~\ref{fig:connection}, left).
Since we assume that $G$ is a torus, we can fix $S_i$'s position in our analysis.
Out of all $N^2$ possible placements of $S_j$ on $G$, there are $\ell$ placements that
result in an intersection that meets the above condition. 
Hence the probability of a \emph{connection} 
between $A_i$ and $A_j$ is $\ell/N^2$ in this scenario.

\item Both $S_i$ and $S_j$ are horizontal and in opposite directions.
In this case, any geometric intersection is enough for a connection to happen
(see Figure~\ref{fig:connection}, right). If both agents move in the same row,
there would be $2\ell$ placements of $S_j$ once $S_i$'s position is fixed that
results in an intersection. The probability of both agents moving in the same 
row is $1/N$, which makes the overall probability $2\ell/N^2$.


\item Both $S_i$ and $S_j$ are horizontal and in the same direction.
In this case, the starting points of $S_i$ and $S_j$ have to be in the same
exact node, which happens with probability $1/N^2$.
\end{enumerate}

Based on the $4$ possible directions of each segment, there are $16$ possible cases for $S_i$
and $S_j$, all of which happens with equal probability and can be reduced
to one of the $3$ cases above. 
The overall probability of a connection between two agents can be bounded as:
\begin{equation}\label{eq:connection}
\frac{\ell}{N^2} \leq \Pr\{\text{Connect}(A_i, A_j, \ell)\} \leq 32\frac{\ell}{N^2} .
\end{equation}
\end{proof}
An immediate consequence of Lemmas \ref{lem:l/n^2-connection} and \ref{lem:strong-overlap} is:
\begin{lemma}\label{lem:connection}
Two agents with strongly overlapping moves have a connection with probability $\Theta(1/N)$.
\end{lemma}
\begin{proof}
For the probability of two strongly overlapping agents connecting we have:
\begin{equation}\label{eq:strong-connection}
\Pr\{\text{Connect}(A_i, A_j)\} = \sum_{x = 1}^{N/2}P(\ell = x)\Pr\{\text{Connect}(A_i, A_j, x)\},
\end{equation}
where $P(\ell = x)$ is the probability that the time duration of the overlap between segments of $A_i$ and
$A_j$ is $x$.
From Lemma~\ref{lem:strong-overlap} we know that $P(\ell \geq N/8) \geq 1/2$. As a result we can rewrite \eqref{eq:strong-connection} as:
\begin{equation*}
\Pr\{\text{Connect}(A_i, A_j)\} \geq \frac{1}{2} \frac{N/8}{N^2} \geq \frac{1}{16N}.
\end{equation*}
The upper bound in \eqref{eq:connection} yields:
\begin{equation*}
\Pr\{\text{Connect}(A_i, A_j)\} \leq \max_{x}\Pr\{\text{Connect}(A_i, A_j, x)\} \leq \frac{16}{N}.
\end{equation*}
\end{proof}

%% file: 032-flood.tex
Given the probability of two agents sharing information during their moves, we are now ready to argue for how information propagates to the entire network. For simplicity, we find an upper bound for the broadcast time $T_B$ (one specific message reaching everyone) and extend it to the flood time $T_F$ (all messages reaching everyone). 
There are two issues that we need to address. First, the positions of an agent are temporally correlated---but fortunately, if sufficiently far apart in time, the positions of agents are independent which will help to simplify our analysis. Second, we need to track agents who have been informed and who have not, and analyze how information spreads from the informed ones to the uninformed ones.

\begin{lemma}\label{lem:independence}
The locations of two agents at $2N$ steps apart are independent of each other.
\end{lemma}

\begin{proof}
Let the sequence of destinations chosen by an agent $A$ be $\langle X_0, X_1, \ldots, X_k \rangle$. Observe that
regardless of what $X_i$ is, every node in the torus (including $X_i$) has the same probability of
being $X_{i+2}$. Also, nodes visited between two consecutive destinations are only dependent on those two
destinations. Let $P_t$ and $P_{t+2N}$ be the position of $A$ at time $t$ and $t + 2N$. Since the agents
move along the shortest path towards their destination, a move is at most $N$ steps long. As a result,
at least two destinations will be visited between time $t$ and $t + 2N$ by $A$. Take the destination immediately after $P_t$, $X_i$, and immediately
before $P_{t+2N}$, $X_j$. As noted above, $P_t$ is only dependent on $X_{i-1}$ and $X_i$, while $P_{t+2N}$ is
only dependent on $X_{j}$ and $X_{j+1}$, where $X_i$ and $X_j$ are distinct positions. This makes $P_t$ and $P_{t + 2N}$ completely independent.
%
%
%
\end{proof}

As mentioned earlier, we track the spread of a single bit of information $b$ among the agents.
We first divide the time span of the whole process into windows of size $6N$ time steps, called \emph{cycles}.
The \rank{k} cycle starts at time $6kN$ and ends at time $6(k+1)N$. Each agent would visit at least
$4$ destinations between time $(6k + 1)N$ and $(6k + 5)N$, which yield at least $3$ moves independent
of other cycles in an agent's trajectory (note the $2N$ margin between selected moves in each cycle,
which guarantees independence). According to Lemma \ref{lem:strong-overlap}, we know that a move
by an agent will strongly overlap with a move by another agent and the starting point of the two moves
are at most $N$ time steps apart.
As a result, for each pair of agents and each cycle,
we can find two strongly overlapping moves independent of other cycles,
which according to Lemma \ref{lem:connection} have a $c/N$ chance of
connection, where $c$ is a constant. This essentially makes collocation events between
a fixed pair of agents in different cycles i.i.d.

We now divide the whole process into consecutive \emph{epochs}.
Each epoch consists of $s_i$ cycles, and starts when we have a set $I_i$ of agents who
know about $b$ (referred to as \emph{informed} agents) and a set $U_i$ of agents who do not (referred to as \emph{uninformed} agents).
An epoch ends when the number of informed agents doubles, or the number of uninformed agents drops to zero.
For the broadcast time of $b$, $T_B$, we can write:
\begin{equation}\label{eq:total-time}
T_B \leq 6N \sum_{i=1}^{\log{M}} s_i.
\end{equation}


We now find the number of cycles needed for that w.h.p.\ each agent in $I_i$ is paired
with a \emph{distinct} agent in $U_i$ during the \rank{i} epoch. By artificially forcing
informed agents to find distinct partners, we only slow down the process
of information spread, and an upper bound found in this manner is valid as
an upper bound for the main problem. The reason we are require distinct partners
for each informed agent is to ensure that connections between different pairs are independent.

The probability of an informed agent $A \in I_i$ connecting to \emph{any} $A' \in U_i$ is as follows
(arrow shows the direction of information exchange):
\begin{align}\label{eq:teach}
\Pr\{\not\exists A' \in U_i: A \rightarrow A'\} &= \Big(1 - \frac{c}{N}\Big)^{s_i|U_i|},  \nonumber\\
\Pr\{\exists A' \in U_i: A \rightarrow A'\} &= 1 - \Big(1 - \frac{c}{N}\Big)^{s_i|U_i|}.
\end{align}

Let there be an arbitrary order for agents in $I_i$ and one for agents in $U_i$.
The first informed agent can match to any of the $|U_i|$ uninformed agents.
After the first matching is done, there will be $|U_i|-1$ potential matches for the second informed agent and so on.
Assuming that there are $q$ pairs at the end of this epoch ($q\leq \min{(|I_i|,|U_i|)}$) and
using Equation~\eqref{eq:teach}, the probability of this happening (i.e., having $q$ pairs of matched informed/uninformed agents) is:
\begin{align*}
\Pr\{\text{$q$-matching}\} &= \prod_{i=0}^{q} \Big(1 - \big( 1 - \frac{c}{N}\big)^{s_i(M-q-i)} \Big)\\
 &\geq \Big(1 - \big( 1 - \frac{c}{N}\big)^{s_i(M/2)} \Big)^{q/2}\\
 &\geq \Big(1 - \exp{\Big(\frac{-s_ic(M/2)}{N}\Big)}\Big)^{q/2}\\
 &\geq 1 - q\exp{\Big(\frac{-s_ic(M/2)}{N}\Big)}.
\end{align*}
For the above to happen w.h.p., for $d > 1$, we need:
\begin{align*}
1 - q\exp{\Big(\frac{-s_ic(M/2)}{N}\Big)} &\geq 1 - \frac{1}{N^d}\\
\exp{\Big( \frac{s_ic(M/2)}{N} \Big)} &\geq qN^d\\
s_i &\geq \frac{2N}{cM}(\log{q} + d \log{N})\\
s_i &> \frac{2dN}{cM}\log{(NM)}.
\end{align*}
The last step is due to the fact that $q< M$. To make sure that
$s_i$ is a non-zero integer and we have at least one cycle, we set:
\begin{equation}\label{eq:c_k}
s_i = \ceil[\Big]{\frac{2dN}{cM}\log{(NM)}}.
\end{equation}
Substituting \eqref{eq:c_k} into \eqref{eq:total-time}, we have:
\begin{align} \label{eq:final_long_walk}
T_B &\leq 6N \sum_{i=1}^{\log{M}}\ceil[\Big]{\frac{2dN}{cM}\log{(NM)}} \\
&= O\Big(N\log{M}\ceil[\Big]{\frac{N}{M}\log{(NM)}}\Big).
\end{align}
Since our bound for $T_B$ works for arbitrarily high
probability ($1-1/N^d$, for a constant $d>0$), it extends to $T_F$ using the Union Bound.
This completes the proof of Theorem~\ref{theorem:main}.

This is a pretty tight bound.
Consider a semi-dense scenario where $M \approx N$, our bound becomes $\tilde{O}(N)$, which nearly meets the trivial lower bound of Equation \eqref{eq:triv_lb}.

Compared to previously mentioned bounds for the \RWP model in \eqref{eq:bound_manhattan} and \eqref{eq:markov_rwp}, our bound is stronger.
For the same movement model, although under slightly different network assumptions, Clementi et al. found the bound of $\tilde{O}(N+N^3/M)$~\cite{ManhattanFlood}, which is mainly due to the
choice of not having a torus as they intended to study the impact of rarely visited areas 
on the total information spread time. Our bound also improves that of \cite{MarkovBound} by a huge margin. This can be due to the fact that their method is a general framework to find an upper bound for $T_F$.
Our version of the \MRWP model assumes that agents complete their
move in one coordinate then start moving in another. As suggested in \cite{MarkovBound},
this assumption increases the probability of connection between two agents,
which in turn leads to a better upper bound for $T_F$ and $T_B$. Further, the \MRWP
model is a better fit for mobility in urban areas it implicitly incorporates the cost of turning during movement.

Depending on the application, one can think of various extensions to our model,
such as an arbitrary transmission radius or random waiting time between two
consecutive moves of an agent. In this case, an approach similar to ours can
be adopted to find a bound for $T_F$ if these three components are available:
(1) suitably sized independent time windows (called \emph{cycle} here), (2) guarantee of a long enough time interval overlap
between segments of moves by two agents and (3) probability of connection between those two segments.

%% file: 041-simulation.tex
\begin{figure*} 
\centering
\begin{tabular}{ccc}
\hspace{-0.15in}
\includegraphics[width=2.4in, height=1.8in]{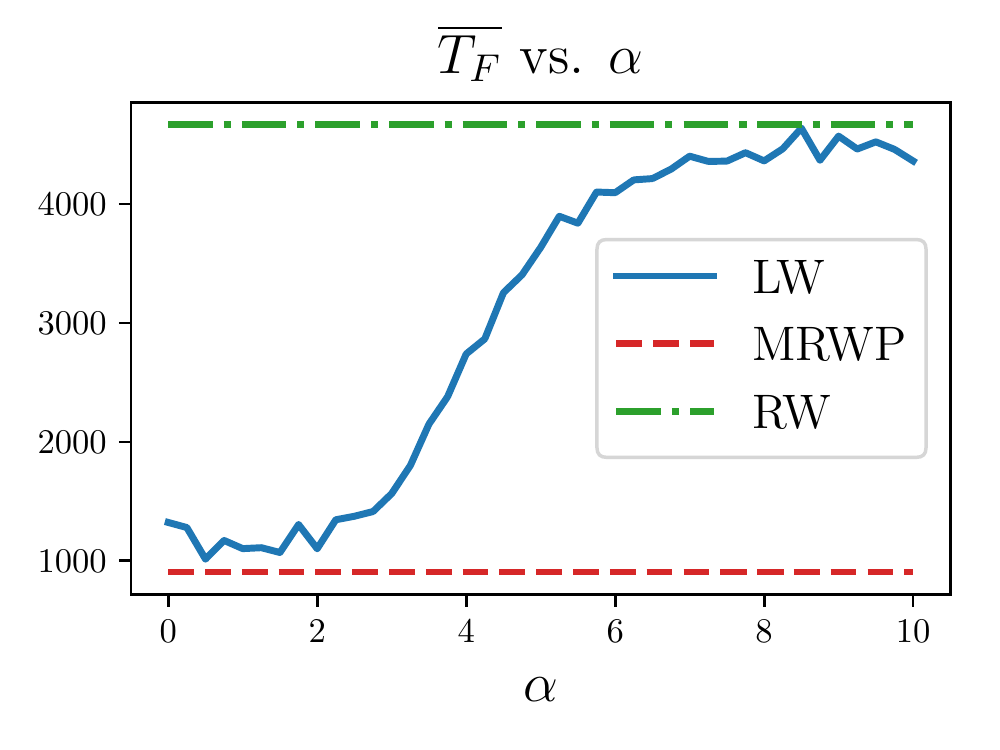}&
\hspace{-0.25in}
\includegraphics[width=2.4in, height=1.8in]{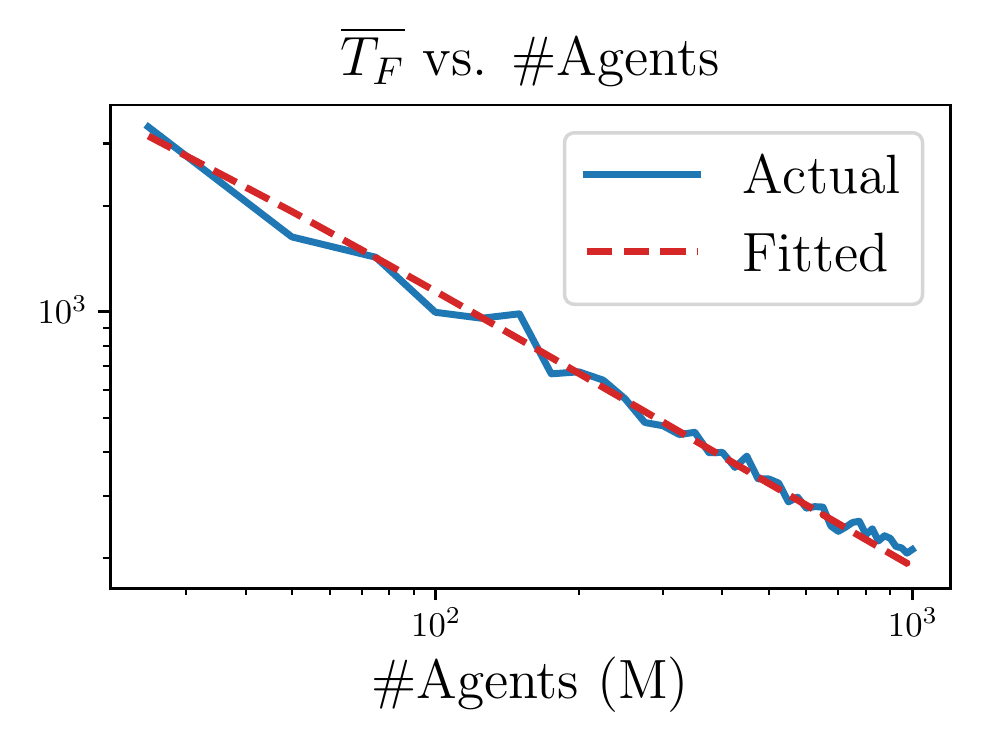}&
\hspace{-0.25in}
\includegraphics[width=2.4in, height=1.8in]{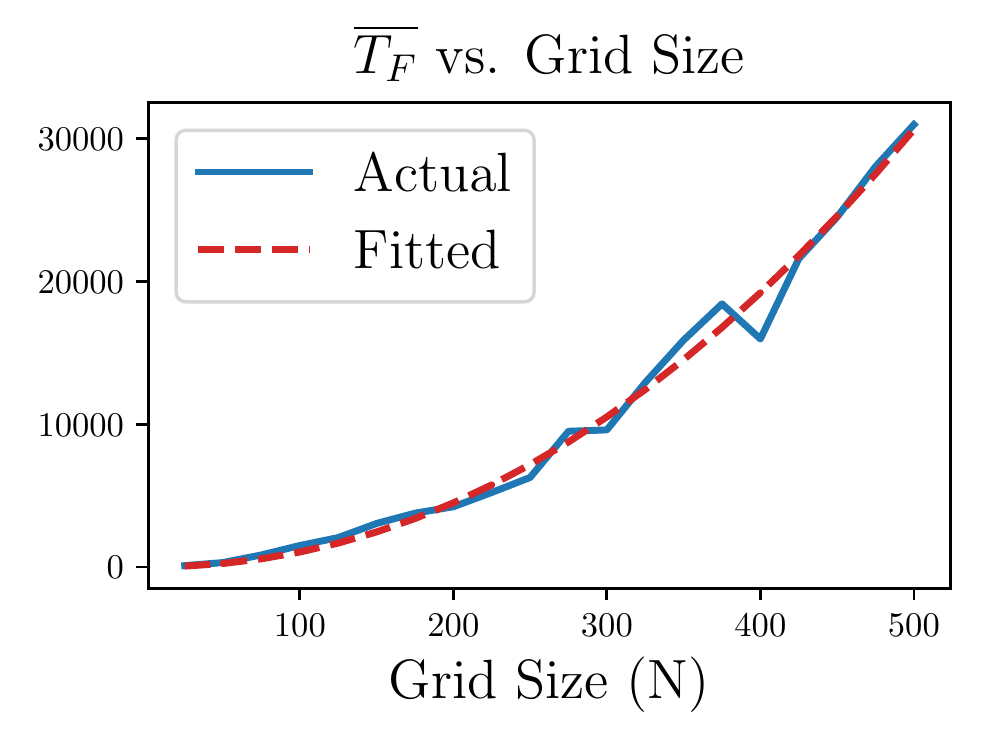}
\end{tabular}
\caption{(left) The sigmoid-like behavior of \LWNoSpace, sandwiched by \RW and \MRWPNoSpace.
(middle, right) Our bound accurately captures the changes in $\overline{T_F}$ as $M$ and $N$
are tweaked while other parameters are fixed.}
\label{fig:simulation}
\end{figure*}

\subsection{Simulated Movements in a Grid}\label{sec:syn-grid}

We simulate the movement of 
agents in a torus-like grid following a \MRWP model, and compare the average
flood times, $\overline{T_F}$, against our bound. First, we fix $N$ to $100$ and find $\overline{T_F}$ for $M = \{
25, 50, \cdots, 1000\}$ by averaging $T_F$ over $100$ realizations. To fit the resulting values to our bound, we use
function $f(M) = \frac{c_1}{M} \log{(c_2 M)} \log{(c_3 M)}$, where each $c_i$ is a positive constant,
accounting for fixed $N$ and constants in our asymptotic analysis. The results of the simulation  along with fitted values are shown
in Figure~\ref{fig:simulation}, middle. Our bound has accurately captured the changes in
$\overline{T_F}$ as the coefficient of determination, $R^2$, is equal to $0.9894$.
Next, using a similar procedure, we fix $M$ to $100$ and find $\overline{T_F}$ for $N = \{25, 50, \cdots, 500\}$. We fit the values to the function $f(N) = c_1 N^2 (\log{N + c_2})$, where again each $c_i$ is a positive constant. The simulation results and the fitted values are depicted in Figure~\ref{fig:simulation}. As expected, we showed good performance here too by yielding an $R^2$ of $0.9894$ (equality between the two $R^2$ values is coincidental).

\subsection{Simulated Movements in Real Networks}\label{sec:syn-bike}

\begin{figure*} 
\centering
\begin{tabular}{ccc}
\vspace{-0.2in}
\hspace{-0.15in}
\includegraphics[width=2.3in, height=2.3in]{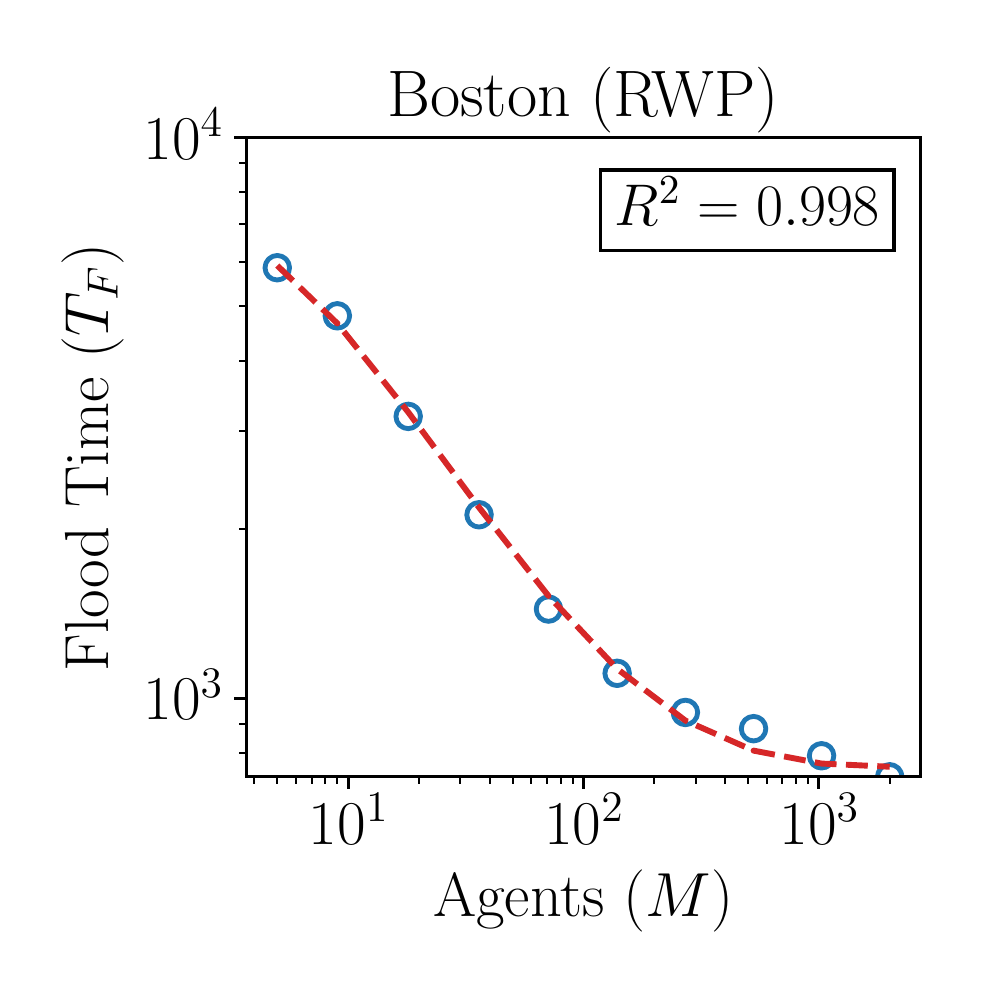}&
\hspace{-0.25in}
\includegraphics[width=2.3in, height=2.3in]{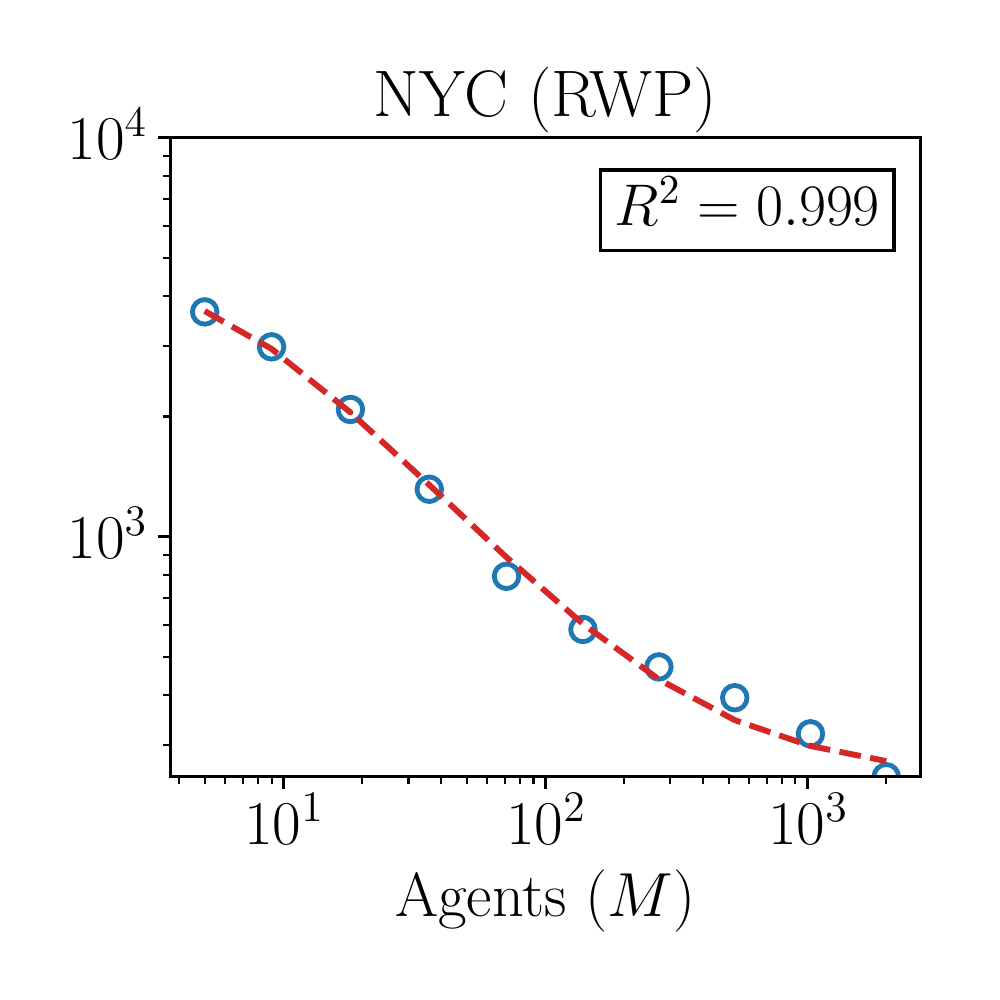}&
\hspace{-0.25in}
\includegraphics[width=2.3in, height=2.3in]{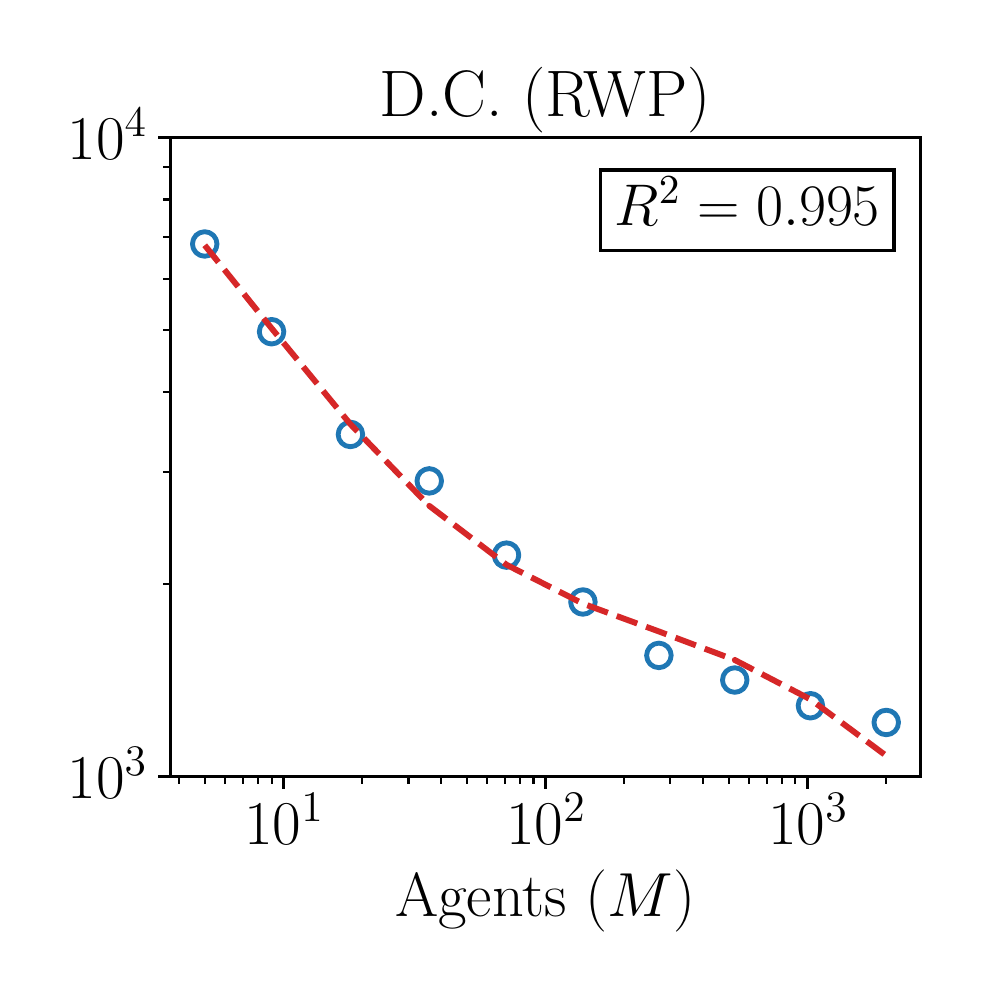}\\
\vspace{-0.15in}
\includegraphics[width=2.3in, height=2.3in]{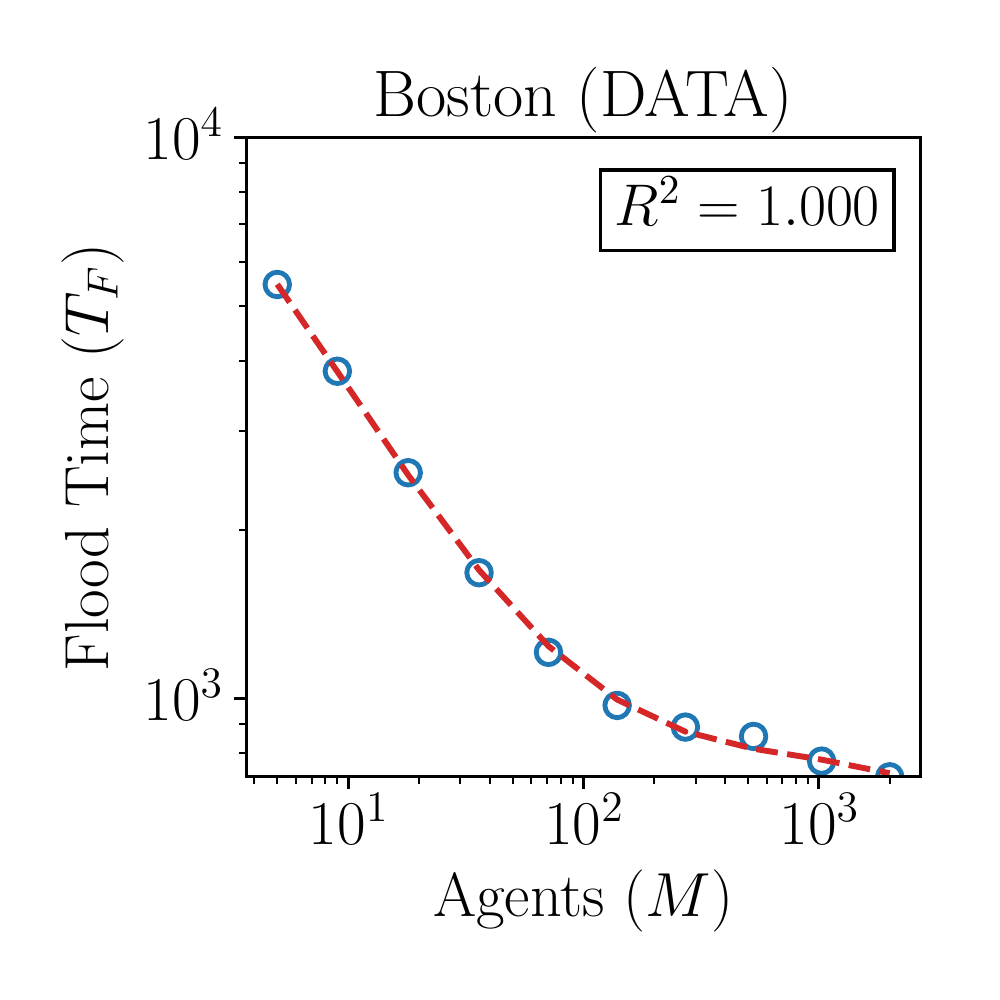}&
\hspace{-0.25in}
\includegraphics[width=2.3in, height=2.3in]{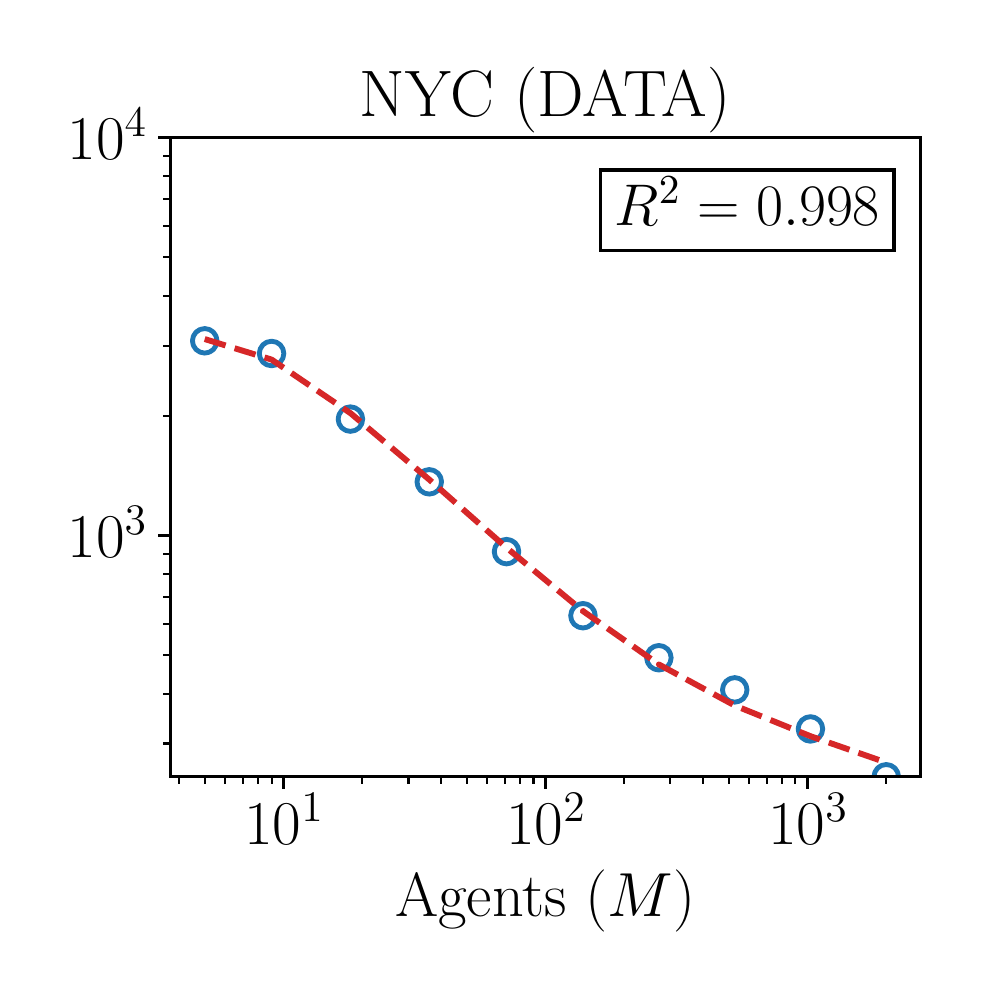}&
\hspace{-0.25in}
\includegraphics[width=2.3in, height=2.3in]{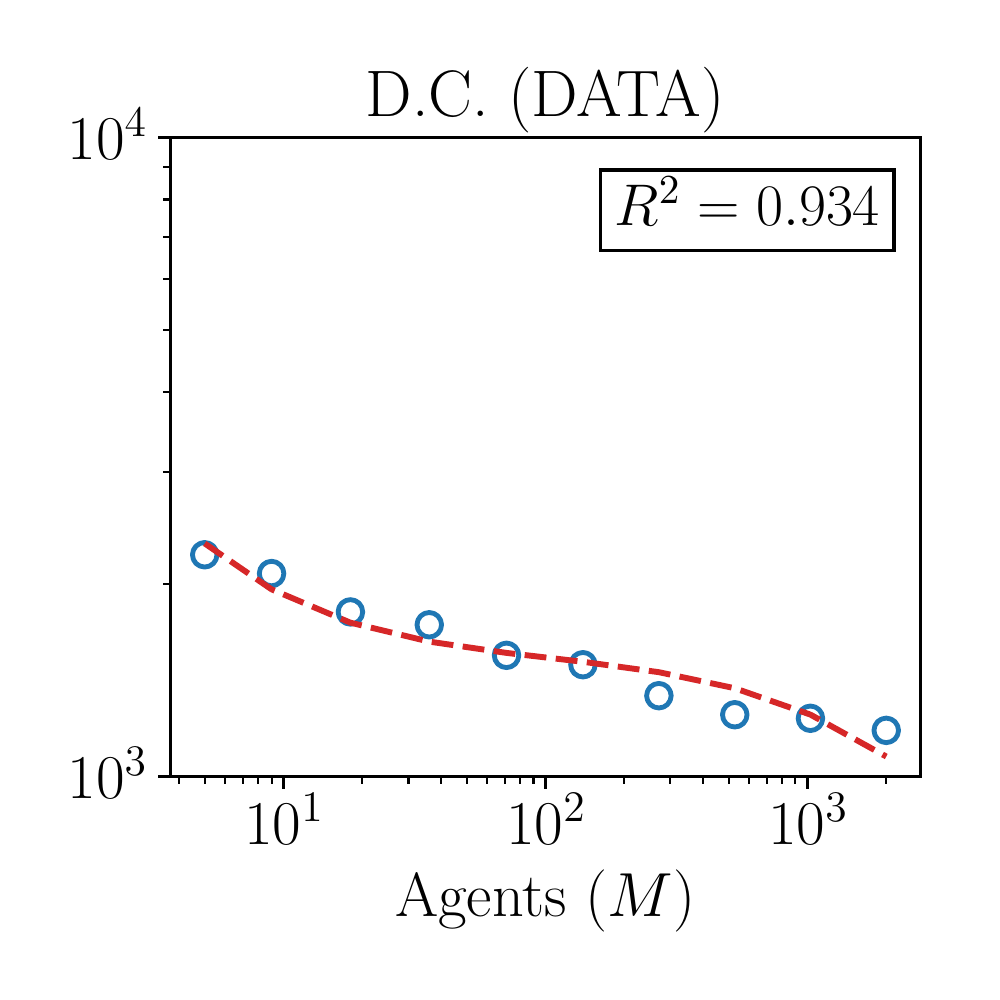}
\end{tabular}
\caption{Actual (empty circles) and fitted (dotted lines) $\overline{T_F}$ for two movement policies in $3$ cities.}
\label{fig:bike_fit}
\end{figure*}

To test our model against non-grid networks,
we use the bike rental records of 3 major US cities~\cite{NYCBikeData,DCBikeData,BostonBikeData}.
Each city has a unique road network and a set of fixed stations,
$\{ S_1, S_2, \cdots, S_k \}$, which are used as the
set of possible destinations each agent can choose from.
The goal here is to test our model against a setting beyond the
grid network and uniformly random selection of destinations.
The data sets include an origin and a destination station for each trip made.
Using these records, we can estimate the probability of choosing a destination $S_j$
given that an agent is currently positioned in station $S_i$, called the
\emph{transition} probability between $S_i$ and $S_j$ and denoted by $P(S_i, S_j)$.
We can also calculate the probability of initiating a trajectory from any
given station, from here on called initiation probability and denoted by $P(S_i)$.
To gradually move our tests away from the theoretical settings, we use the following movement models throughout our simulations:
\begin{enumerate}
\item Similar to Section~\ref{sec:syn-grid}, we select each station in the sequence of stations visited by an agent uniformly at random. This is equivalent to the \RWP model and denoted by \emph{RWP} in this experiment.
\item Next, we use the calculated $P(S_i)$ and $P(S_i, S_j)$ values to build synthetic trajectories. We call this model \emph{DATA} throughout this experiment.
\end{enumerate}

In each experiment, after selecting a sequence of stations visited by each agent, we find the shortest paths between consecutive stations using Routino~\cite{Routino} and
OpenStreetMap~\cite{OpenStreetMap} extracts.
Two agents will \emph{connect}, if at any time $t$ they are closer than
$100$ meters from each other.
Finally, for each of these experiments, we iterate over $10$ values of $M$ between $5$ and $2000$, and report the average flood time ($\overline{T_F}$) by aggregating over 25 realizations.

%

Using a function similar to Section~\ref{sec:syn-grid}, we can fit the simulation results to our bound. Figure~\ref{fig:bike_fit} shows the actual and fitted values for the 3 cities, along with the $R^2$ value of the fitting. In these simulations, our bound closely approximates the flood time in the simulations, even when the movement policy used is data-specific rather than the \RWP model. The $R^2$ values care compared in Table~\ref{table:bike_sim_r2}.
Across all settings, we achieve $>0.93$ ($5$ of them $>0.99$), which shows the flexibility of our model to variations of network and movement policy. Note that here the network was a real-world road network, and far from a torus.

There can be many different factors contributing to the flood time in networks as complicated
as urban maps, which are beyond the scope of this study. Here, we tried to
explore the limits of our model's prediction capabilities by tweaking the settings of experiment
in a controlled manner. Further investigation in the effects
of structural properties of road networks, and different distributions of
frequent origins and destinations on the flood time is needed to fully understand the process of information dissemination
by human mobility in real road networks.
\begin{table}
\centering
\caption{Fitting score, $R^2$, for all movement policies in all cities.}
\label{table:bike_sim_r2}
\begin{tabular}{|c | c c |}
 \hline
 \textbf{City} & \textbf{RWP} & \textbf{DATA} \\
 \hline
 Boston  & $0.998$   & $1.000$  \\
New York City  & $0.999$ & $0.998$  \\
Washington, D.C.  & $0.995$   & $0.934$  \\
 \hline
\end{tabular}
\end{table}

%% file: 042-real.tex
\begin{figure}[!tbh]
  \includegraphics[width=.75\linewidth]{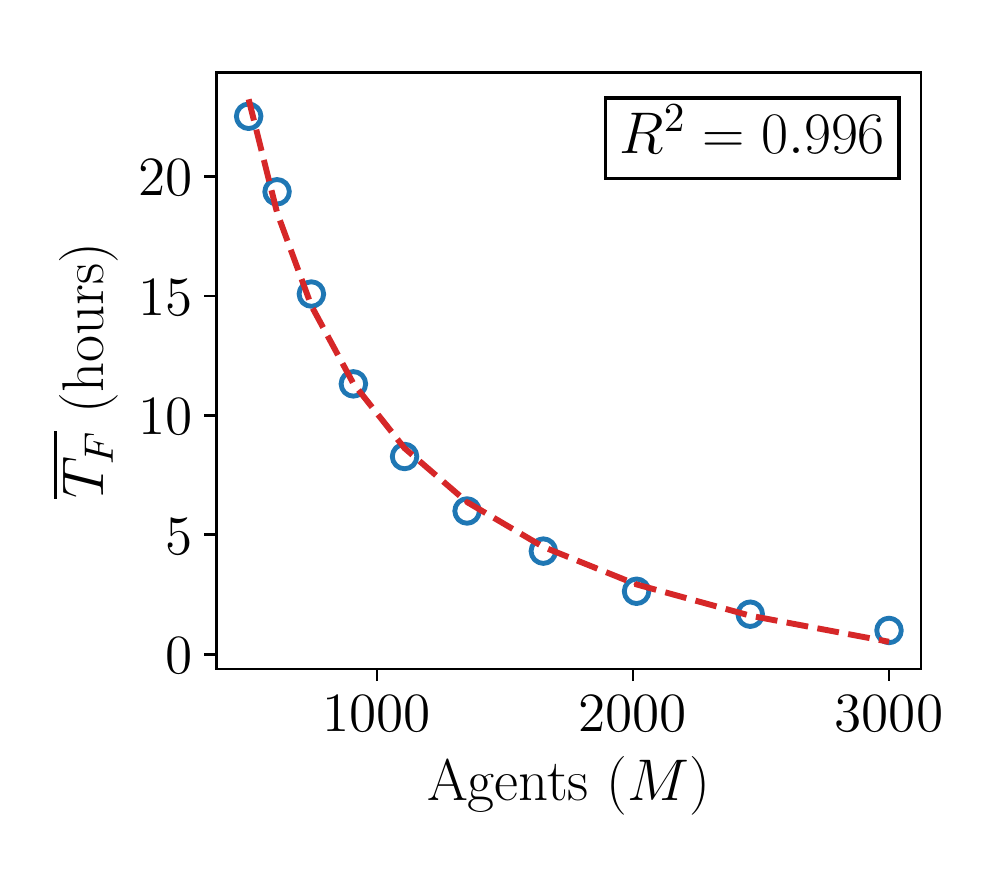}
\caption{Actual (empty circles) and fitted (dotted lines) $\overline{T_F}$ for Shenzhen.}
\label{fig:shenzhen}
\end{figure}

Next, we try to fit our model to real-world GPS traces.
Ideally, one may want to experiment on personal trajectories, as the behavior
of a \emph{single} moving agent is best understood by looking at individuals'
mobility traces. However, due to the sensitivity of such data, large and high-quality
data sets containing personal mobility traces are extremely rare.
As a substitute, we can study the mobility of shared vehicles, as we did in the previous section.
Here, we study GPS traces of taxi cabs in the city of Shenzhen in China~\cite{JiaxinVehicleNetwork}.
Over the course of $24$ hours, the location of $9386$ taxis are sampled every $1.01$ minutes.
We set the \emph{transmission radius} to $100$ meters and, for simplicity, assume that
connections can happen only on sampled points in time. To generate different numbers of
moving agents ($M$), we have to subsample from the set of all taxis. Since these trajectories
are fixed, we cannot extend them in the event of having no information flood. Hence, we filter out
those taxis that meet less than $100$ distinct taxis during the whole 24 hours, and $3905$ taxis will remain.
We iterate over $10$ different values of $M$ between $500$ and $3000$, each time
finding the flood time in hours. We average the results of $25$ realizations for each $M$ and report it.
The results are shown in Figure~\ref{fig:shenzhen}. We have followed the same procedure to fit
the simulation values to our bound. The resulting \emph{fitted} line is drawn in Figure~\ref{fig:shenzhen},
achieving an $R^2$ value of $0.996$. This shows that our model is capable of predicting
flood times for real-world scenarios to some degree. It is worth noting that the real-world experiments
did not have significant fluctuations in the flood time value and, similar to controlled experiments in the two sections before,
shows a smooth behavior, even with only hundreds of moving agents in some cases.

%% file: 050-levy.tex
\begin{figure}[!tbh]
  \includegraphics[width=\linewidth]{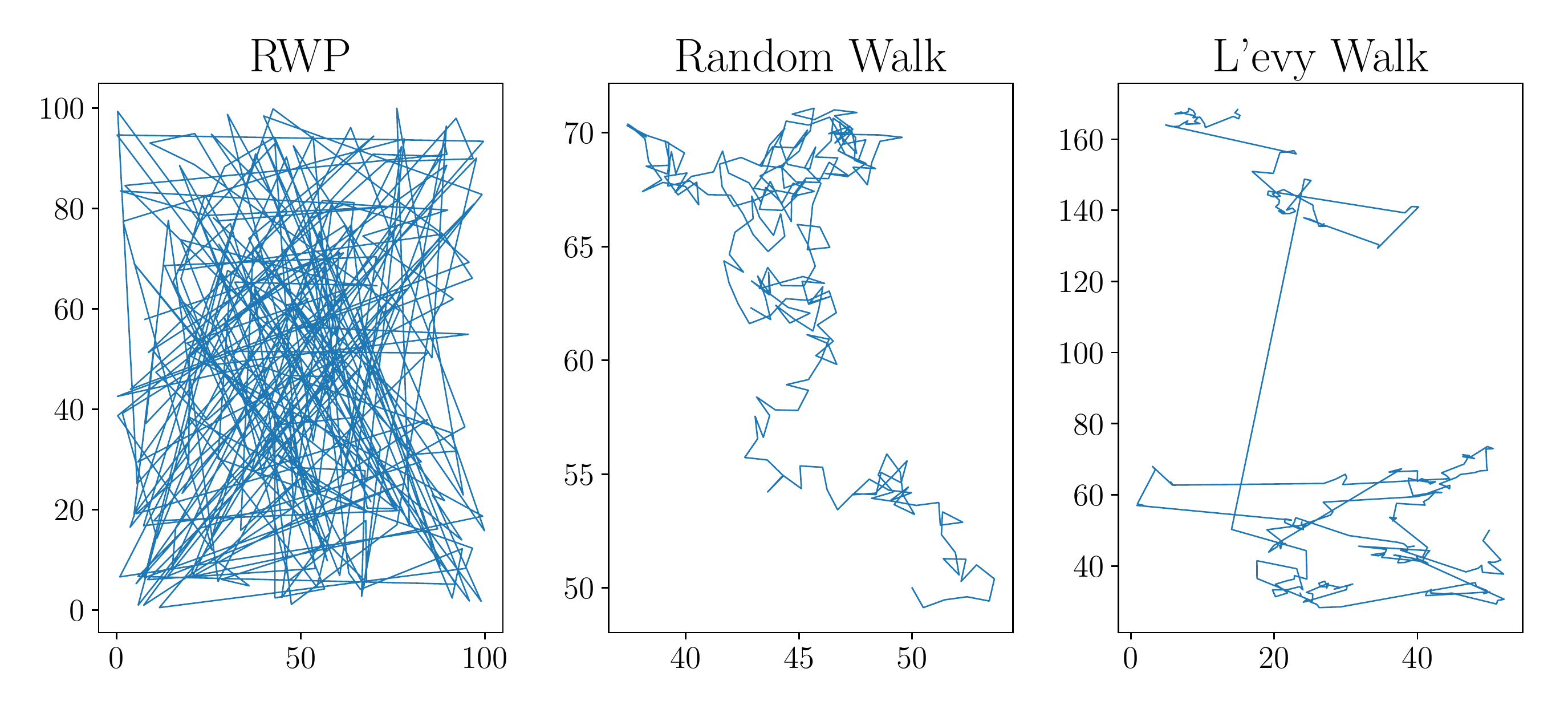}
\caption{A \RWP walker (left), a random walker (middle), and a \Levy walker ($\alpha=2$) after 250 steps.}
\label{fig:walk_samples}
\end{figure}

Compared to other mobility models, the \LW if far less studied. Formally, in a \LWNoSpace, given a constant $\alpha>0$, an agent
positioned at its \rank{i} destination, $X = X_i$, chooses node $Y$ as its next destination, $X_{i+1}$, with the following probability:
\begin{equation} \label{eq:levy}
\Pr\{X_{i+1} = Y \mid X_i = X\} = \frac{1}{Z||Y - X||^\alpha},
\end{equation}
where $Z$ is the normalizing factor and $||X - Y||$ is the distance between
nodes $X$ and $Y$, such as \emph{Manhattan Distance} in a grid,
\emph{Euclidean Distance} in the 2D plane or \emph{Graph Shortest Path Distance} in any given network.
Figure~\ref{fig:walk_samples} compares a Random, a \Levy and a \RWP walker, simulated for 250 steps.
Notice that a \RWP walker tends to take big steps and cover a vast area in the grid,
while a Random walker is concentrated to a small area around its initial position.
A \Levy walker shows a mixture of the two behaviors. It roams around in a small area most of the time, but
occasionally makes a long move to a different region in the grid.

To compare how information propagates in the three movement models, we have to first 
observe that \RW (or \textit{Brownian motion}) and \RWP can be thought of as two extreme ends of
the spectrum of all possible \LWNoSpace s.
In \eqref{eq:levy}, setting $\alpha$ to $0$ (and applying the corresponding $Z$ value)
yields a constant probability regardless of
the distance between $X$ and $Y$, similar to \RWPNoSpace.
On the other hand, given any time limit $T$, we can make $\alpha$ high enough so that w.h.p.
no agent selects a destination more than one unit distance away at any time $t \leq T$,
effectively forcing them to follow \RWNoSpace.

Figure \ref{fig:simulation}, left, shows the simulated results for
the average flood time, denoted by $\overline{T_F}$, of a system where agents are moving by \RWNoSpace, \MRWP or \LW model with different values of $\alpha$. Both $N$ and $M$ are set to $100$, $\alpha$ goes from $0$ to $10$ in $0.1$ increments, and each point is created by aggregating the results of 100 realizations.
Additionally, with our newly discovered bound for the \MRWP model, the bounds for $\overline{T_F}$ in \RW and \MRWP have gotten very close. We now have a reason to believe that any future bound for \LWNoSpace\ should be close to either of the bounds for these two movement models.
And since their bounds are close, it is worth investigating whether or not a careful interpolation of the bounds for \RW and \MRWP
is a good predictor of how a \Levy Walker moves in a network.

%% file: 060-conclusion.tex
Thanks to ever-present portable devices, there has been a growing interest in a better understanding of
mobile networks (also called \emph{vehicular} networks), where autonomous agents move independently and
are capable of carrying and transmitting information. We studied the case of $M$ agents moving
in an $N \times N$ torus.

We made a new improvement to the flood time bound for \MRWP model, $T_F = \tilde{O}(N\ceil[\big]{N/M})$,
 that is tight for a wide range of problem settings.
To the best of our knowledge, this bound is stronger than all previous bounds found for this movement model.
Through extensive experiments, we showed that our bound can accurately predict flood time for a wide variety of
simulated and real-world settings.

Lastly, given the shrinking difference between the bounds for \RW and \RWPNoSpace, and the fact that
\LW behaves in between the former two movement models, it is now worth investigating whether
a careful interpolation of \RW and \RWP can describe \LW accurately enough.
Finding theoretical bounds for \LW can be a valuable future
work that further expands our knowledge of the relation
between these three movement models and ultimately of human mobility.